\documentclass[10pt]{blois07}
\usepackage{graphicx}
\usepackage{bm}
\usepackage{cite,./mcite}
\usepackage{wrapfig,rotating}
\usepackage{epstopdf}
\usepackage{amssymb,amsmath,array}
\begin{document}
\title{Rapidity gap survival in the black-disk 
regime\footnote{\hspace{.3em} Proceedings of EDS07, DESY, 
Hamburg, 21--25 May 2007. 
A preliminary version of this report was published in the proceedings 
of DIS07, Munich, 16--20 Apr.~2007, arXiv:0708.3106 [hep-ph].}
\footnote{\hspace{.3em} Notice: 
Authored by Jefferson Science Associates, LLC under U.S.\ DOE
Contract No.~DE-AC05-06OR23177. The U.S.\ Government retains a
non-exclusive, paid-up, irrevocable, world-wide license to publish or
reproduce this manuscript for U.S.\ Government purposes.}}
\author{L.~Frankfurt$^1$, C.E.~Hyde$^2$, 
M.~Strikman$^3$, C.~Weiss$^4$}
%
%\vspace{.3cm}\\
%
\institute{$^1$ School of Physics and Astronomy, Tel Aviv University, 
Tel Aviv, Israel\\
$^2$ Old Dominion University, Norfolk, VA 23529, USA, and \\
\hspace{1ex} Laboratoire de Physique Corpusculaire, Universit\'e Blaise 
Pascal, 63177 Aubi\`ere, France\\
$^3$ Department of Physics, Pennsylvania State University,
University Park, PA 16802, USA \\
$^4$ Theory Center, Jefferson Lab, Newport News, VA 23606, USA}

\maketitle

\begin{abstract}
We summarize how the approach to the black--disk regime (BDR)
of strong interactions at TeV energies influences rapidity gap survival
in exclusive hard diffraction $pp \rightarrow p + H + p \; 
(H = \textrm{dijet}, \bar Q Q, \textrm{Higgs})$.
Employing a recently developed partonic description of such proces\-ses,
we discuss (a) the suppression of diffraction 
at small impact parameters by soft spectator interactions in the BDR;
(b) further suppression by inelastic interactions of hard spectator 
partons in the BDR; (c) effects of correlations between hard and soft 
interactions, as suggested by various models of proton structure 
(color fluctuations, spatial correlations of partons). Hard spectator 
interactions in the BDR substantially reduce the rapidity gap survival 
probability at LHC energies compared to previously reported estimates.
\end{abstract}

\section{Introduction}
At high energies strong interactions enter a regime in which 
cross sections are comparable to the ``geometric size'' of 
the hadrons, and unitarity becomes an essential feature of 
the dynamics. By analogy with quantum--mechanical scattering
from a black disk, in which particles with impact
parameters $b < R_{\textrm{disk}}$ experience inelastic interactions 
with unit probability, this is known as the black--disk
regime (BDR). The approach to the BDR is well--known
in soft interactions, where it generally can be attributed to the 
``complexity'' of the hadronic wave functions. It is seen \textit{e.g.}\
in phenomenological parametrizations of the $pp$ elastic scattering 
amplitude, whose profile function $\Gamma(b)$ approaches unity at 
$b = 0$ for energies $\sqrt{s} \gtrsim 2 \, \textrm{TeV}$. More recently
it was realized that the BDR is attained also in hard processes 
described by QCD, due to the increase of the gluon density in the proton 
at small $x$. Theoretically, this phenomenon can be studied in the 
scattering of a small--size color dipole ($d \sim 1/Q$) from the proton. 
Numerical studies show that  at $\sqrt{s}\sim$ few TeV 
%energies 
the dipole--proton
interaction is close to ``black'' up to $Q^{2} \sim 
\textrm{several 10 GeV}^2$ \cite{Frankfurt:2005mc}. This fact has 
numerous implications for the dynamics of $pp$ collisions at the LHC,
where multiple hard interactions are commonplace. For example, 
it predicts dramatic changes in the multiplicities and $p_T$ spectra 
of forward particles in central $pp$ collisions
compared to extrapolations of the Tevatron data \cite{Frankfurt:2003td}.
Absorption and energy loss of leading partons by inelastic interactions
in the BDR can also account for the pattern of forward pion production 
in $d$--$Au$ collisions at STAR \cite{Frankfurt:2007rn}.

Particularly interesting is the question what the approach to the 
BDR implies for exclusive hard diffractive scattering,
$pp \rightarrow p + H + p$. In such processes a high--mass system
($H = \textrm{dijet}, \bar QQ, \textrm{Higgs})$ is produced in
a hard process involving exchange of two gluons between the protons.
At the same time, the spectator systems must interact in a way such 
as not to produce additional particles. This restricts the set
of possible trajectories in configuration space and results in a 
suppression of the cross section compared to non-diffractive events. 
For soft spectator interactions this suppression is measured by the
so--called rapidity gap survival (RGS) probability. Important questions
are (a) what role the BDR plays in traditional soft--interaction RGS; 
(b) how the physical picture of RGS is modified by hard spectator 
interactions in the BDR at LHC energies; (c) how fluctuations of the 
strength of the $pp$ interaction related to inelastic diffraction
influence RGS in hard diffractive processes; (d) how possible correlations
between hard and soft interactions affect RGS.

These questions can be addressed in a recently proposed partonic
description of exclusive diffraction \cite{Frankfurt:2006jp}, 
based on Gribov's parton picture of high--energy hadron--hadron scattering. 
Questions (a) and (b) can be studied within this framework in a practically 
model--independent way. They require only basic information about the
strength of hard and soft interactions and their impact parameter
dependence, which is either known experimentally or can be obtained
from reasonably safe extrapolations of existing data to higher energies.
Questions (c) and (d) require more detailed assumptions about correlations
in the partonic wavefunction of the proton, which relate to less understood 
features of the $pp$ interaction at high energies. We can address them
by implementing within the approach of Ref.~\cite{Frankfurt:2006jp}
specific dynamical models of nucleon structure (color fluctuations,
transverse correlations between partons). Our studies of these questions
are of exploratory nature.
\section{Black--disk regime in soft spectator interactions}
%
% FIGURE
%
\begin{figure}
\begin{tabular}{lcl}
\parbox[c]{0.25\columnwidth}{
\includegraphics[width=0.25\columnwidth]{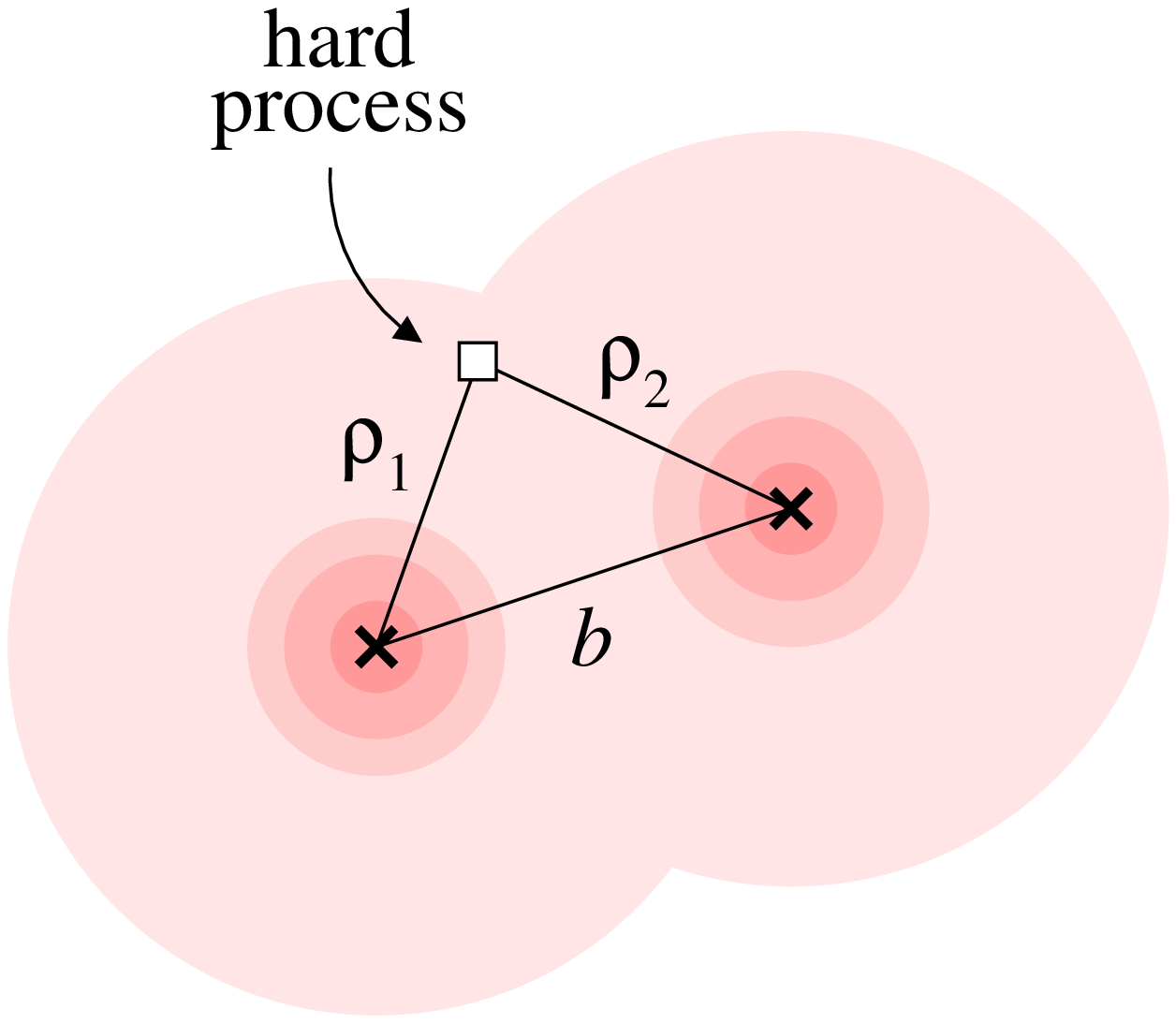}
} 
& \hspace{0.1\columnwidth} &
\parbox[c]{0.55\columnwidth}{
\includegraphics[width=0.55\columnwidth]{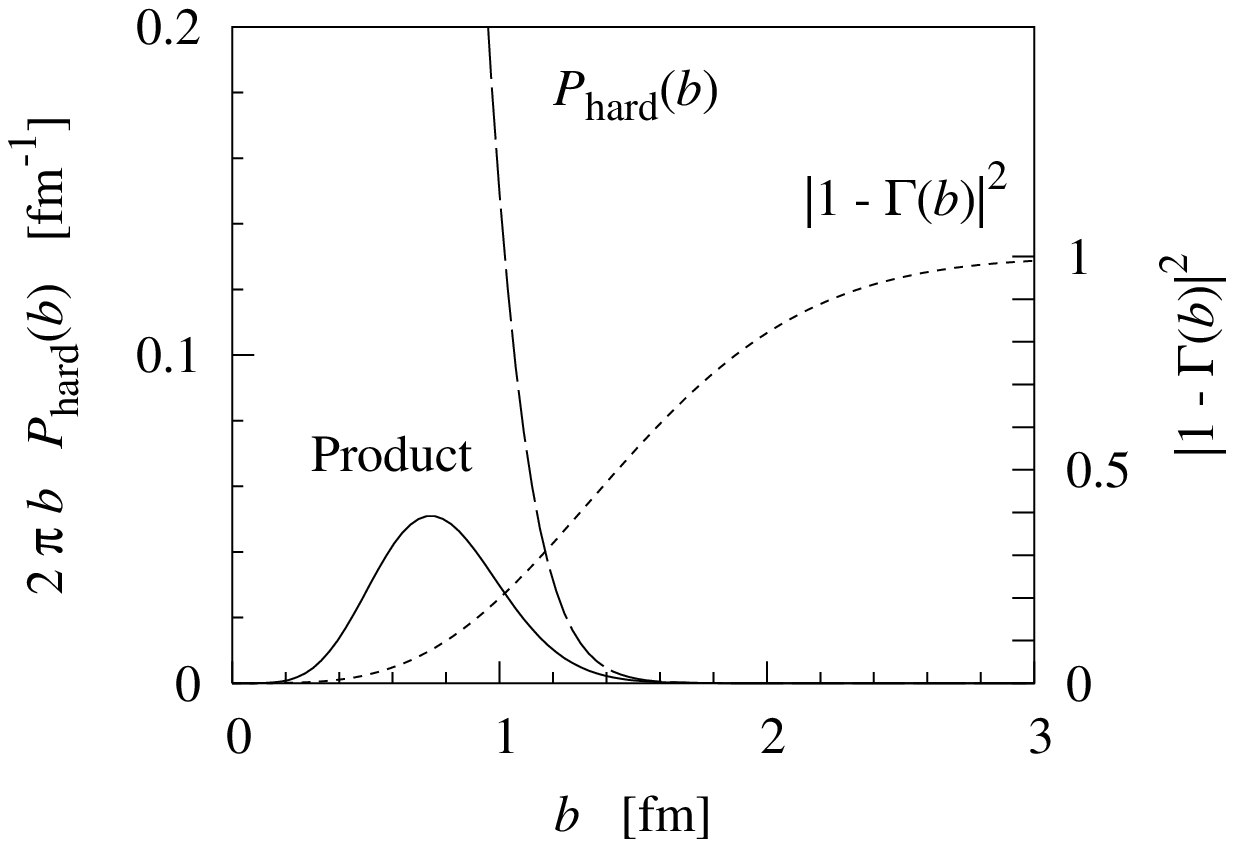}}
\\[-3ex]
{\small (a)} & & {\small (b)}
\end{tabular}
\caption[]{(a) Transverse geometry of hard diffractive $pp$ scattering.
(b) Dashed line: Probability for hard scattering process 
$P_{\textrm{hard}} (\bm{b})$ as function of the $pp$ impact parameter, $b$.
Dotted line: Probability for no inelastic interactions between
the protons, $|1 - \Gamma (\bm{b})|^2$.
Solid line: Product $P_{\textrm{hard}} (\bm{b}) |1 - \Gamma (\bm{b})|^2$.
The RGS probability (\ref{survb}) is given by the area under this curve.
The results shown are for Higgs production at the LHC
($\sqrt{s} = 14\, \textrm{TeV}, M_H \sim 100\, \textrm{GeV}$).
(We point out that the distributions shown in Fig.~8 of 
Ref.~\cite{Frankfurt:2006jp} correspond 
to a gluon $t$--slope $B_g = 4\, \textrm{GeV}^{-2}$, 
not $B_g = 3.24\, \textrm{GeV}^{-2}$ as stated in the caption. 
The plot here shows the correct distributions
for $B_g = 3.24\, \textrm{GeV}^{-2}$.)}
\label{Fig:rgs}
\end{figure}
A simple ``geometric'' picture of RGS is obtained in the approximation 
where hard and soft interactions are considered to be 
independent \cite{Frankfurt:2006jp}. The hard two--gluon
exchange process can be regarded as happening locally in space--time 
on the typical scale of soft interactions. In the impact parameter
representation (see Fig.~\ref{Fig:rgs}a) the RGS probability can be
expressed as
\begin{equation}
S^2 \;\; = \;\; 
\int d^2 b \; P_{\textrm{hard}} (\bm{b}) \; |1 - \Gamma (\bm{b})|^2 .
\label{survb}
\end{equation}
Here $P_{\textrm{hard}} (\bm{b})$ is the probability for two hard gluons 
from the protons to collide in the same space--time point, given by
the overlap integral of the squared transverse spatial distributions
of gluons in the colliding protons, normalized to 
$\int d^2 b \; P_{\textrm{hard}} (\bm{b}) = 1$
(see Fig.~\ref{Fig:rgs}b). 
The function $|1 - \Gamma (\bm{b})|^2$ is the probability for the
two protons not to interact inelastically in a collision with
impact parameter $b$. The approach to the BDR in $pp$ scattering
at energies $\sqrt{s} \gtrsim 2\, \textrm{TeV}$
implies that this probability is practically zero at small 
impact parameters, and becomes significant only for 
$b \gtrsim 1\, \textrm{fm}$ (see Fig.~\ref{Fig:rgs}b).
This eliminates the contribution from small
impact parameters in the integral (\ref{survb}) 
(see Fig.~\ref{Fig:rgs}b) and determines the value of the 
RGS probability to be $S^2 \ll 1$.
One sees that the approach to the BDR in soft interactions
plays an essential role in RGS at high energies.
\section{Black--disk regime in hard spectator interactions}
\label{sec:BDR_hard}
\begin{figure}
\begin{tabular}{lcl}
\parbox[c]{0.2\columnwidth}{
\includegraphics[width=0.2\columnwidth]{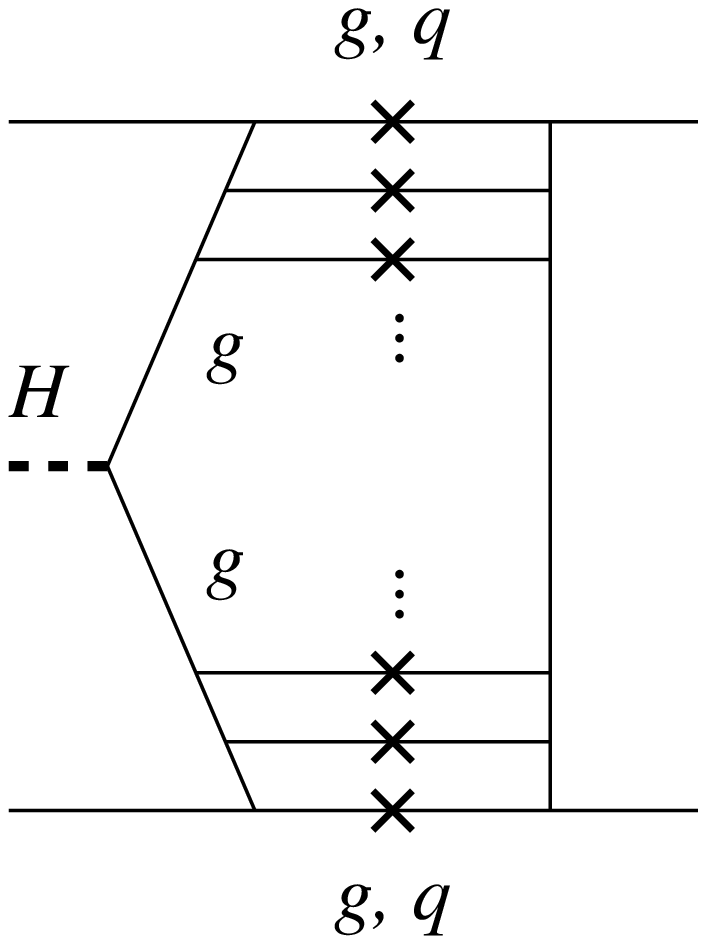}}
&
\hspace{.2\columnwidth}
&
\parbox[c]{0.44\columnwidth}{
\includegraphics[width=0.44\columnwidth]{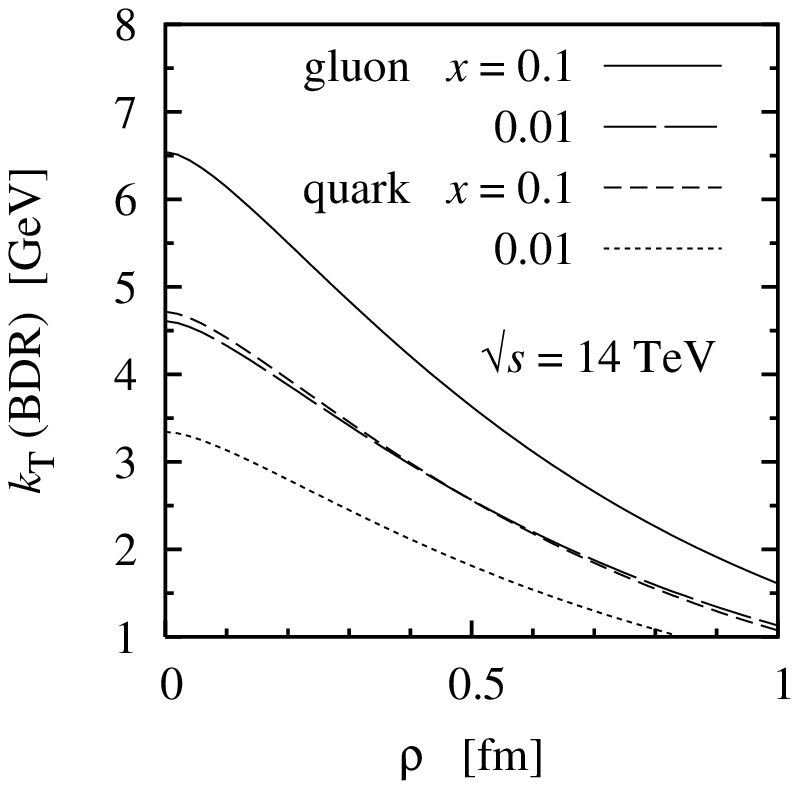}}
\\[-4ex]
{\small (a)} & & {\small (b)}
\end{tabular}
\caption[]{(a) Absorption of parent partons by interactions in the BDR. 
The crosses denote absorptive interactions with 
small-x gluons in the other proton.(b) The critical 
transverse momentum, $k_T(\textrm{BDR})$, 
below which partons are absorbed with high probability
($|\Gamma^{\textrm{parton-proton}}| > 0.5$), as a function of the
parton--proton impact parameter, $\rho = \rho_{1, 2}$.}
\label{Fig:hardscreen}
\end{figure}

At LHC energies even highly virtual partons ($k^2 \sim \textrm{few GeV}^2$)
with $x \gtrsim 10^{-2}$ experience ``black'' interactions with the 
small--$x$ gluons in the other proton. This new effect causes
an additional suppression of diffractive scattering which is not 
included in the traditional RGS probability \cite{Frankfurt:2006jp}. 
One mechanism by which this happens is the absorption of ``parent''
partons in the QCD evolution leading up to the
hard scattering process (see Fig.~\ref{Fig:hardscreen}a). 
Specifically, in Higgs production at the LHC the gluons producing 
the Higgs have momentum fractions 
$x_{1, 2} \sim M_H / \sqrt{s} \sim 10^{-2}$; their ``parent'' partons
in the evolution (quarks and gluons)
typically have momentum fractions of the order 
$x \sim 10^{-1}$ and transverse momenta $k_T^2 \sim \textrm{few GeV}^2$.
Quantitative studies of the BDR in the dipole picture show
that at the LHC energy such partons are absorbed with near--unit 
probability if their impact parameters with the other proton 
are $\rho_{1, 2} \lesssim 1 \, \textrm{fm}$ 
(see Fig.~\ref{Fig:hardscreen}b). For proton--proton impact parameters 
$b < 1 \, \textrm{fm}$ about $90\%$ of the strength in 
$P_{\textrm{hard}}(b)$ comes from parton--proton impact parameters 
$\rho_{1, 2} < 1 \, \textrm{fm}$ (\textit{cf}.\ Fig.~\ref{Fig:rgs}a),
so that this effect practically eliminates diffraction at 
$b < 1 \, \textrm{fm}$. Since $b < 1 \, \textrm{fm}$ accounts for 2/3 
of the cross section [see Eq.~(\ref{survb}) and Fig.~\ref{Fig:rgs}b)],
and the remaining contributions at $b > 1 \, \textrm{fm}$ are also 
reduced by absorption, we estimate that inelastic interactions
of hard spectators in the BDR reduce the RGS probability at LHC
energies to about 20\% of its soft--interaction value. 

%
% FIGURE
%
In the above argument one must also allow for the possibility of 
trajectories with no gluon emission. Mathematically, they correspond 
to the Sudakov form factor--suppressed $\delta(1 - x)$--term in the 
evolution kernel. While such trajectories are not affected by absorption,
their contributions are small both because of the Sudakov suppression,
and because they effectively probe the gluon density at the soft input 
scale, $Q_0^2 \sim 1 \, \textrm{GeV}^2$. The probability for a gluon 
not to emit a gluon when evolving from virtuality $Q_0^2$ to $Q^2$, 
is given by the square of the Sudakov form factor, 
\begin{equation}
C\;\; = \;\; \left[S_G(Q^2/Q_0^2)\right]^2 
\;\; = \;\; \exp\left( -\frac{3\alpha_s}{\pi} \ln^2 \frac{Q^2}{Q_0^2} \right) .
\end{equation}
At the same time, each of the parton densities in the trajectory 
without emissions is suppressed compared to those with emissions
by a factor $g(x, Q^2)/ g(x, Q_0^2)$, where $Q^2 \sim 4 \, \textrm{GeV}^2$. 
The overall relative suppression of trajectories without emission
is thus by a factor
\begin{equation}
R\;\; = \;\; C^2 \left[ \frac{g(x, Q^2)}{g(x, Q_0^2)} \right]^2 
\;\; \sim \;\; \frac{1}{10} .
\end{equation}
Although this contribution is suppressed, it is comparable to that
of average trajectories with emissions because the latter
are strongly suppressed by the absorption effect described above.
Combining the two, we find an overall suppression factor 
of the order $\sim 0.3$. In order to make more accurate estimates one 
obviously would need to take into account fluctuations in the number
of emissions more carefully. In particular, trajectories on which 
only one of the partons did not emit gluons, which come with a 
suppression factor of $\sqrt{R}$, may give significant contributions.

The approach to the BDR in hard spectator interactions described here
``pushes'' diffractive $pp$ scattering to even larger impact parameters
than are allowed by soft--interaction RGS (except for the 
Sudakov--suppressed contribution discussed in the previous paragraph).
This should manifest itself in a shift of the final--state proton 
transverse momentum distribution to smaller values, which could be 
observed in $p_T$--dependent measurements of diffraction at the LHC.

The estimates reported here are based on the assumption that DGLAP 
evolution reasonably well describes the gluon density down to 
$x \sim 10^{-6}$; the quantitative details (but not the basic picture) 
may change if small--$x$ resummation corrections were to significantly 
modify the gluon density at such values of $x$ 
(see Ref.~\cite{Ciafaloni:2007gf} and references therein). 
The effect of hard spectator interactions described here
is substantially weaker at the Tevatron energy.

\section{Color fluctuations in the colliding protons}
In the approximation where hard and soft interactions in the diffractive
process are considered to be independent, the RGS probability can be
expressed through the $pp$ elastic scattering amplitude, and effects
of inelastic diffraction do not enter into consideration, see 
Ref.~\cite{Frankfurt:2006jp} and the discussion above. It is important
to investigate how accurate this approximation is in practice, and how 
correlations between hard and soft interactions modify the picture.
Such correlations generally arise from correlations between partons
in the wave functions of the colliding protons, which can be caused
by several physical mechanisms, see Ref.~\cite{Frankfurt:2006jp}
for a discussion. Here we focus on one mechanism which is closely related
to the presence of inelastic diffraction channels, namely fluctuations 
of the size of the interacting configurations (color 
fluctuations). Our study of this effect here is of exploratory nature;
details will be reported in a forthcoming publication.

\begin{figure}
\begin{tabular}{lcl}
\parbox[c]{0.28\columnwidth}{
\includegraphics[width=0.28\columnwidth]{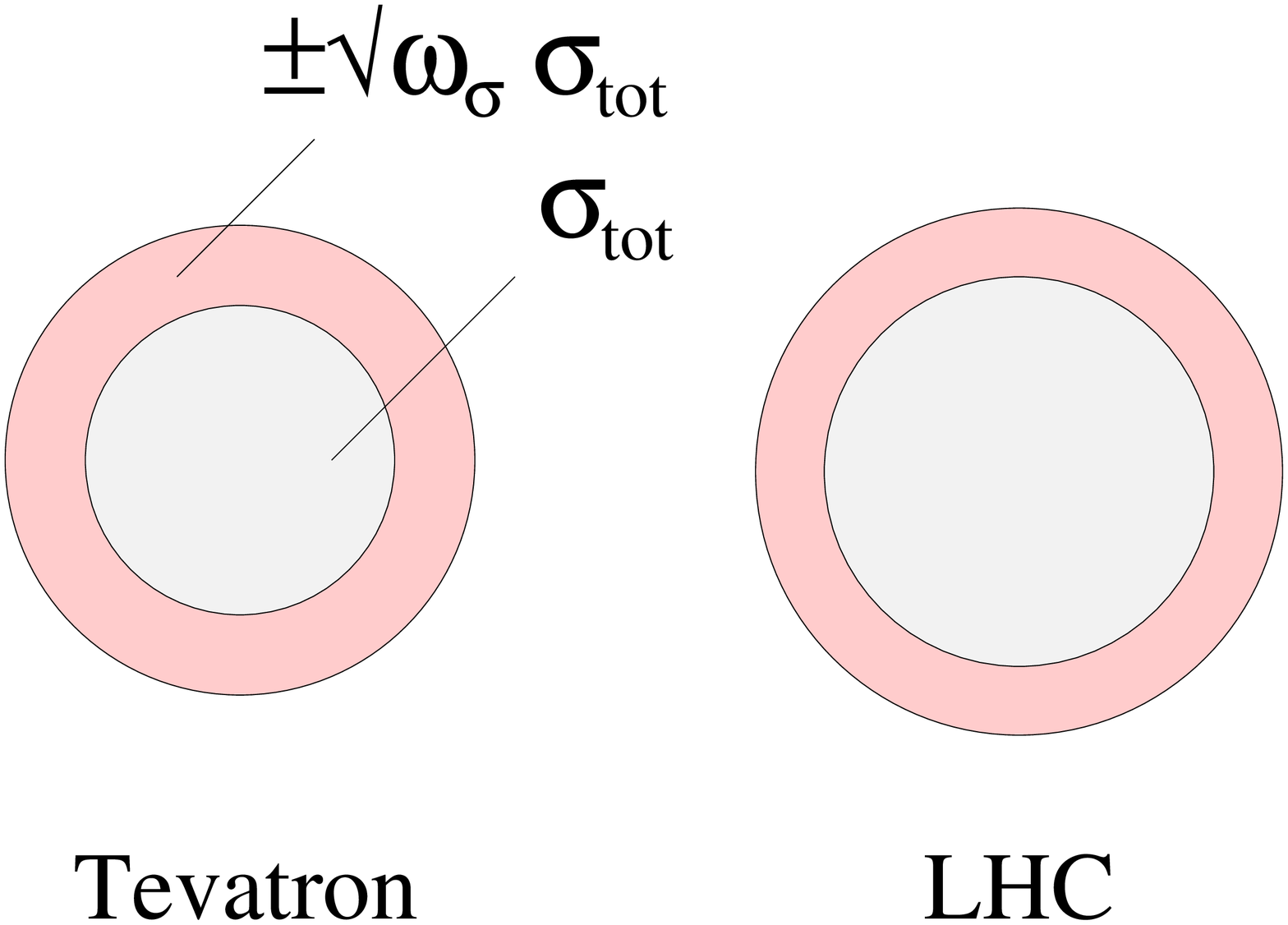}
} 
& \hspace{0.01\columnwidth} &
\parbox[c]{0.62\columnwidth}{
\includegraphics[width=0.62\columnwidth]{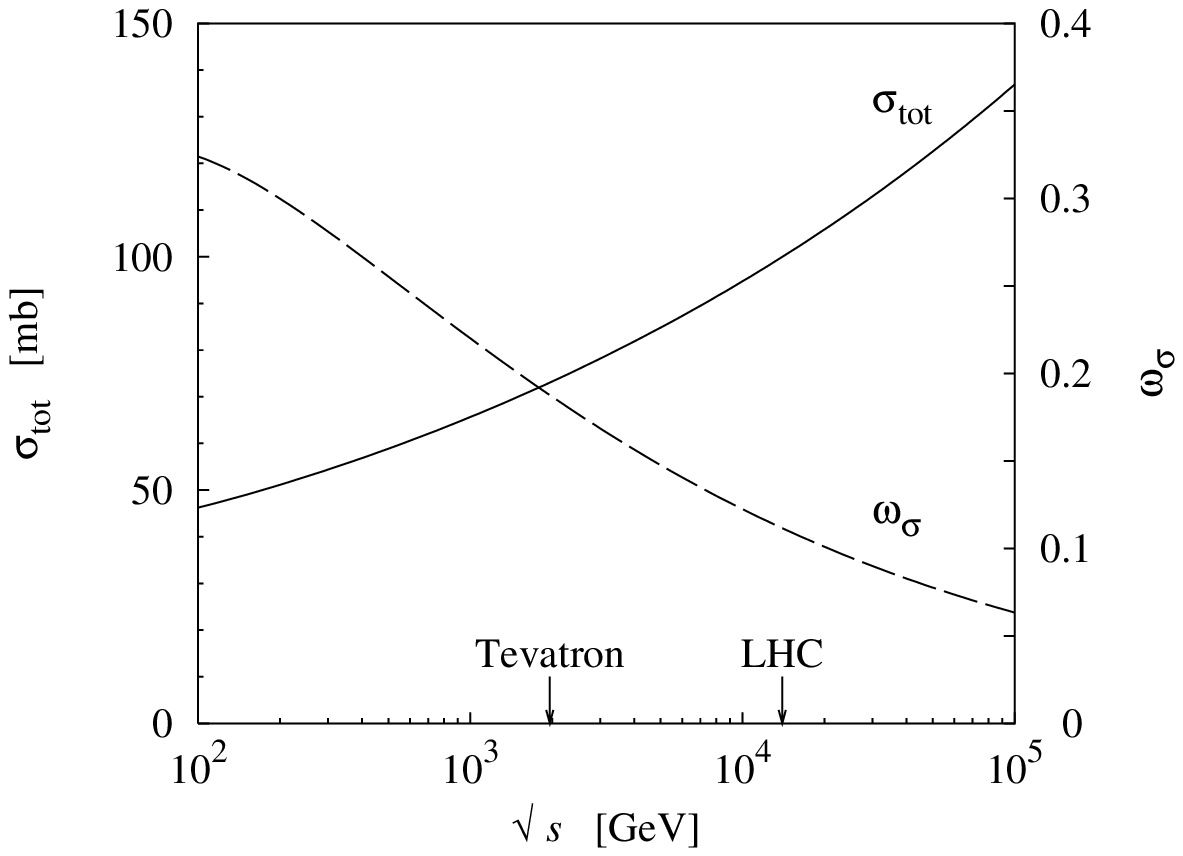}}
\\[-3ex]
{\small (a)} & & {\small (b)}
\end{tabular}
\caption[]{(a)~Graphical representation of the cross section 
distributions in diffraction at the Tevatron and LHC energy. 
The area of the inner and outer disk at given energy
is proportional to $(1 \pm \sqrt{\omega}_\sigma ) 
\langle\sigma\rangle$, \textit{i.e.}, the average area 
represents the average cross section $\langle\sigma\rangle = 
\sigma_{\text{tot}}$, the difference (ring) the range of the 
fluctuations $\pm \sqrt{\omega}_\sigma \langle\sigma\rangle$.
(b)~The $s$--dependence of the total cross section $\sigma_{\textrm{tot}}$ 
(left $y$--axis) and the dispersion $\omega_\sigma$ (right $y$--axis), 
as predicted by a Regge--based parametrization of $\sigma_{\textrm{tot}}$
\cite{Donnachie:1992ny} and a parametrization of the inelastic 
diffractive cross section $d\sigma_{\textrm{inel}}/dt |_{t = 0}$, 
measured up to the Tevatron energy \cite{Goulianos}.
The weak energy dependence of the width of the ring in figure~(a) 
reflects the slow variation of the diffractive cross section with energy.}
\label{Fig:stot_om}
\end{figure}

The basic idea is that in diffractive high--energy scattering the colliding 
hadrons can be regarded as a superposition of configurations of different 
size, which are ``frozen'' during the time of the interaction.
In the well--known approach of Good and Walker \cite{Good:1960ba} 
this is implemented by expanding the incident hadron state in 
eigenstates of the $T$--matrix of the same quantum numbers.
A more general formulation uses the concept of the cross section
distribution, $P(\sigma)$, which can be interpreted as the probability for 
the hadron to scatter in a configuration with given cross 
section, with $\int d\sigma \, P(\sigma ) = 1$ \cite{Blaettel:1993ah}. 
It is defined such that its average reproduces the total cross section,
\begin{equation}
\langle \sigma \rangle \;\; \equiv \;\; 
\int d\sigma \; \sigma \; P(\sigma) \;\; = \;\; \sigma_{\textrm{tot}} ,
\end{equation}
while its dispersion coincides with the ratio of the differential 
cross sections for inelastic ($pp \rightarrow p + X$) and elastic
($pp \rightarrow p + p$) diffraction at $t = 0$ \cite{Miettinen:1978jb},
\begin{equation}
\omega_{\sigma} \;\; \equiv \;\; 
\frac{ \langle\sigma^2\rangle - \langle\sigma\rangle^2}
{\langle\sigma\rangle^2}
\;\; = \;\; 
\left. \frac{d\sigma_{\text{inel}}}{dt} \right/ \left.
\frac{d\sigma_{\text{el}}}{dt} \right|_{t=0} .
\end{equation}
The dispersion and the third moment of $P(\sigma)$ have been extracted 
from analysis of the $pp$ and $pd$ data up to 
$s \approx 8\times 10^{2} \, \text{GeV}^2$; at higher energies 
the shape of the distribution is not well known. Extrapolation of a 
parametrization of the Tevatron data \cite{Goulianos} suggests that 
between the Tevatron and LHC energy $\omega_\sigma$ should drop by a factor 
$\sim 2$, while at the same time the total cross section is expected to 
grow, indicating that the relative magnitude of fluctuations decreases 
with increasing energy (see Figs.~\ref{Fig:stot_om}a and b). Generally, 
one should expect that the different configurations in diffractive scattering 
are characterized by a different parton density. In hard diffractive 
processes $pp \rightarrow p + H + p$ this effect would then lead to a 
modification of the ``independent interaction'' result for the RGS 
probability, Eq.~(\ref{survb}).
%
% FIGURE
%

The theoretical description of the role of cross section fluctuations 
in hard diffraction is a complex problem, which requires detailed
assumptions about the proton's partonic wave function. Here we aim
only for a simple phenomenological estimate, which illustrates
the sign and order--of--magnitude of the effect, as well as its 
energy dependence. Our basic assumption is that the strength 
of interaction in a given configuration is proportional to the 
transverse area occupied by color charges. To implement this
idea, we start from the cross section distribution $P(\sigma )$ 
at fixed--target energies ($s \lesssim 8\times 10^{2} \, \textrm{GeV}^2$), 
which can be related to the fluctuations of the size 
of the basic ``valence quark'' configuration in the proton wave function
and is known well from the available data \cite{Blaettel:1993ah}. 
We then assume that
\begin{itemize}
\item[(a)] 
The parton density is correlated with the parameter 
$\sigma$ characterizing the size of the interacting configuration. 
One simple scenario is to assume that the parton density changes 
with the size of the configuration only through its dependence on the 
normalization scale, $\mu^2 \propto R_{\textrm{config}}^{-2} \propto \sigma$.
This is analogous to the model of the EMC effect of Ref.~\cite{Close:1983tn}, 
and leads to a simple scaling relation for the $\sigma$--dependent
gluon density,
\begin{equation}
g(x,Q^2 \, | \sigma ) \;\; = \;\; g(x, \xi Q^2),
\hspace{3em}
\xi (Q^2) \;\; \equiv \;\; 
\left( \sigma / \langle \sigma \rangle 
\right)^{\alpha_s (Q_0^2) / \alpha_s (Q^2)} ,
\label{g_sigma}
\end{equation}
where $Q_0^2 \sim 1 \, \textrm{GeV}^2$. In Higgs boson production 
one expects $Q^2 \approx 4 \, \textrm{GeV}^2$, and $x = M_H/\sqrt{s} 
= 0.007 \, \textrm{(LHC)}, 0.05 \, \textrm{(Tevatron)}$ 
with $M_H = 100 \, \textrm{GeV}$. An alternative scenario 
--- the constituent quark picture --- will be discussed below.
\item[(b)] The size distribution in soft high--energy interactions
is correlated with the parameter $\sigma$ characterizing the
valence quark configuration. As a minimal model we assume that
soft interactions in a configuration with given 
$\sigma$ is described by a profile function of the form
\begin{equation}
\Gamma (\bm{b}, s \, | \sigma ) \;\; = \;\; 
\exp\left[-\frac{2\pi b^2}{\sigma_{\text{tot}}(s, \sigma)}\right] ,
\hspace{2em}
\textrm{with}
\hspace{1em}
\sigma_{\text{tot}}(s, \sigma) \;\; = \;\; \alpha (s) + \beta (s) 
\frac{\sigma}{\langle \sigma \rangle} ,
\label{Gamma_sigma}
\end{equation}
in which the parameters $\alpha(s)$ and $\beta(s)$ are chosen such 
as to reproduce the average cross section and dispersion of the 
high--energy cross section distribution (see Fig.~\ref{Fig:stot_om}a and b)
when averaging over the (given) $\sigma$ distribution
$P(\sigma)$. Note that the profile in Eq.~(\ref{Gamma_sigma}) approaches 
the black--disk limit at $b \rightarrow 0$, and that the average 
elastic profile $\langle \Gamma (\bm{b}, s \, | \sigma ) \rangle$ 
obtained in this way is very close to that found in 
the standard phenomenological parametrizations of
the $pp$ elastic and total cross section data. More sophisticated 
parametrizations could easily be constructed
but would not change our qualitative conclusions.
\end{itemize}
With assumptions (a) and (b) we can estimate the effect of 
color fluctuations in the protons in hard diffraction in a simple way.
In the presence of correlations between the parton density and the
strength of soft interactions, the RGS probability is now given by
\begin{equation}
S^2_{\text{corr}} \;\; = \;\; 
\int d^2 b \; \left\langle P_{\textrm{hard}}(\bm{b} \, | \sigma)
\;\; |1 - \Gamma (\bm{b} , s \, | \sigma)|^2
\phantom{\frac{0}{0}} \hspace{-.3em}
\right\rangle ,
\label{S2_corr}
\end{equation}
where $P_{\textrm{hard}}(\bm{b} \, | \sigma)$ is the normalized 
impact parameter distribution for the hard process obtained with 
the $\sigma$--dependent gluon density Eq.~(\ref{g_sigma}),
and $\langle \ldots \rangle$ denotes the average over the
$\sigma$ distribution. This should be compared to the RGS probability 
without correlations,
\begin{equation}
S^2_{\text{uncorr}} \;\; = \;\; 
\int d^2 b \; \left\langle P_{\textrm{hard}}(\bm{b} \, | \sigma)
\phantom{\frac{0}{0}} \hspace{-.3em} \right\rangle 
\left\langle
|1 - \Gamma (\bm{b} , s \, | \sigma)|^2
\phantom{\frac{0}{0}} \hspace{-.3em}
\right\rangle ,
\label{S2_uncorr}
\end{equation}
which corresponds to the expression obtained previously in the
approximation of independent hard and soft interactions,
Eq.~(\ref{survb}), if we identify the functions there with
the average distributions.\footnote{Note that there are small differences 
between the functional forms of the $\sigma$--averaged distributions 
in Eq.~(\ref{S2_uncorr}) and the original ($\sigma$--independent) 
distributions used previously in evaluating Eq.~(\ref{survb}).
This is only the result of imperfect modeling of the $\sigma$--dependent 
distributions and immaterial for the physical correlation effect
discussed here.} For a quantitative estimate, we first consider 
fluctuations of the interacting configurations in only one of the 
colliding protons, leaving the other protons unchanged.
In this case we obtain
\begin{equation}
\frac{S^2_{\text{corr}} - S^2_{\text{uncorr}}}{S^2_{\text{uncorr}}}
\;\; = \;\;
-0.15 \hspace{2em} \textrm{at} \hspace{1em} 
\sqrt{s} = \; 2 \, \textrm{TeV} 
\hspace{1em} \textrm{(Tevatron).} 
\label{S2_conn}
\end{equation}
If one could consider the fluctuation effect as a small correction,
the total effect would be additive and thus proportional to the
number of protons, \textit{i.e.}, Eq.~(\ref{S2_conn}) would have to be 
multiplied by 4, corresponding to the two protons in both the amplitude 
and the complex-conjugate amplitude in the cross section.
While the magnitude of the correction Eq.~(\ref{S2_conn}) does 
not really justify such additive treatment, we can at least
to conclude that the overall effect from correlations in this model
should be a reduction of the RGS probability by $\sim 1/2$. 
Note that the sign of the correlation effect simply reflects the fact 
that smaller configurations, which have higher transparency and thus 
larger survival probability, have a lower density of small--$x$ partons
in model adopted here.

Our treatment of color fluctuations here assumes that the basic
picture of independent hard and soft interactions in RGS is still valid, 
and that the fluctuations can be incorporated by way of an ``external''
parameter controlling the size of the interacting configurations. 
As explained above (Sec.~\ref{sec:BDR_hard}) and in 
Ref.~\cite{Frankfurt:2006jp}, this assumption breaks down at the LHC energy, 
where hard spectator interactions approach the BDR. The correction 
described here thus should be valid at RHIC and Tevatron energies
but not at the LHC. In particular, this can be seen in the fact that the 
correlation effect of Eq.~(\ref{S2_conn}) is obtained from modification 
of the impact parameter distribution of hard diffraction 
at $b \lesssim 1\, \textrm{fm}$, where we expect hard spectator
interactions to be ``black'' at the LHC, see 
Sec.~\ref{sec:BDR_hard}. Thus, corrections from inelastic 
diffractive channels of the kind discussed here play a minor role 
at the LHC energy. This is a welcome conclusion, as it means that
our predictions for the RGS probability at the LHC are not substantially
modified by such corrections.
%\footnote{The suppression of 
%inelastic diffraction in the BDR reflects a general physical phenomenon.
%Diffractive dissociation requires different absorption of the various 
%components of the incident wave. ``Black'' interactions (total absorption)
%give rise only to elastic diffraction \cite{Good:1960ba}.}

The numerical estimate of correlation effects reported here was
obtained with the assumption that the gluon density in the interacting 
configurations scales with the size of the configuration as 
in Eq.~(\ref{g_sigma}). Physically, this corresponds to the 
assumption that the valence quark configuration in the proton
acts coherently as source of the gluon field, and that there are
no other physical scales in the proton besides the size of that 
configuration. This is clearly an extreme scenario 
and does not take into account the physical
scales generated by the non-perturbative vacuum structure of QCD.
An alternative scenario would be a constituent quark picture, in which 
the normalization scale of the gluon density is determined by the ``size'' 
of the constituent quark (related to the spontaneous breaking of chiral
symmetry) and not related to the size of the multi--quark configuration 
in the nucleon. For this picture the relation between the 
gluon density and the size of the interacting configuration
would be very different from Eq.~(\ref{g_sigma}). It leads to a 
different kind of correlation between hard and soft 
interactions, see Ref.~\cite{Frankfurt:2006jp} and 
Section~\ref{sec:corr} below.
\section{Transverse spatial correlations between partons}
\label{sec:corr}
%
% FIGURE
%
\begin{wrapfigure}{l}{0.4\columnwidth}
\centerline{\includegraphics[width=0.36\columnwidth]{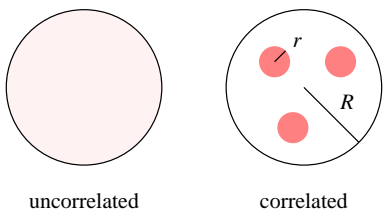}}
\caption[]{Transverse parton correlations.}
\label{Fig:corr}
\end{wrapfigure}
The partonic approach to RGS of Ref.~\cite{Frankfurt:2006jp} also
allows one to incorporate effects of correlations in the partonic
wavefunction of the protons. They can lead to correlations between
hard and soft interactions in diffraction, which substantially
modify the picture of RGS compared to the independent interaction
approximation. The analysis of the CDF data on $p\bar{p}$ collisions 
with multiple hard processes indicate the presence of substantial transverse 
correlations between partons with $x\gtrsim 0.1$ \cite{Frankfurt:2005mc}. 
Such correlations naturally arise in a constituent quark picture 
of the nucleon with $r_q \ll R$ (see Fig.~\ref{Fig:corr}). 
It is interesting that the observed enhancement of the cross 
section due to correlations seems to require $r_q/R\sim 1/3$,
which is the ratio suggested by the instanton vacuum model of 
chiral symmetry breaking (see Ref.~\cite{Diakonov:2002fq} for
a review). Such correlations modify the picture of RGS in hard 
diffractive $pp$ scattering compared to the independent interaction 
approximation in two ways \cite{Frankfurt:2006jp}. 
On one hand, with correlations inelastic interactions between
spectators are much more likely in configurations in which two 
large--$x$ partons collide in a hard process than in average 
configurations, reducing the RGS probability compared to the
uncorrelated case. On the other hand, the ``lumpiness'' implies
that there is generally a higher chance for the remaining spectator 
system not to interact inelastically compared to the mean--field
approximation. A quantitative treatment of correlations in RGS,
incorporating both effects, remains an outstanding problem.

\section{Summary}
The approach to the BDR at high energies profoundly influences
the physics of RGS in exclusive diffractive scattering.
The onset of the BDR in soft spectator interactions at 
$\sqrt{s} \gtrsim 2 \, \textrm{TeV}$ eliminates diffractive scattering
at small impact parameters and determines the basic order--of--magnitude 
of the RGS probability at the Tevatron and LHC. At LHC energies, 
the BDR in hard spectator interactions pushes diffractive scattering
to even larger impact parameters and further reduces the RGS probability
by a factor of 3 (likely more, 4--5), implying that $S^2 < 0.01$,
much smaller than initial estimates reported in the literature,
see Ref.~\cite{Frankfurt:2006jp} and references therein.
At the same time, this effect reduces the relative importance of
color fluctuations related to inelastic diffraction, making our
theoretical predictions of the RGS probability more robust.
At the Tevatron energy, we have seen that color fluctuations lower 
the RGS probability compared to the approximation of independent hard
and soft interactions. The simple model estimate presented in 
Sec.~\ref{sec:corr} suggests reduction by a factor of the order 1/2; 
however, more refined estimates are certainly needed. 
Finally, spatial correlations between partons are likely to modify the picture 
of RGS both at the Tevatron and the LHC energy; a detailed study
of this effect would be of principal as well as of considerable 
practical interest.

The total RGS probability is an ``integral'' quantity, which 
combines contributions from very different trajectories of the
interacting $pp$ system. It is also difficult to determine experimentally, 
as its extraction requires precise knowledge of the cross section of the 
hard scattering process (gluon GPD, effective virtualities, etc.).
Much more detailed tests of the diffractive reaction mechanism
can be performed by studying the transverse momentum dependence
of the diffractive cross section, which can be interpreted without
knowledge of the hard scattering process. In particular, the predicted
onset of the BDR in hard interactions between the Tevatron and LHC
energy (Sec.~\ref{sec:BDR_hard}) should cause substantial narrowing of 
the $p_T$ distribution, which could be observed 
experimentally. At RHIC and Tevatron energies, the correlation effects
described in Sec.~\ref{sec:corr} imply that the $p_T$ distribution
is narrower than predicted by the independent interaction approximation,
allowing one to test this picture experimentally. This underscores the 
importance of planned transverse momentum--dependent measurements of 
diffraction at RHIC and LHC.

\begin{footnotesize}
%

% ----------------------------------------------------------------------------

\end{footnotesize}

\end{document}